\documentclass[journal]{IEEEtran}
\ifCLASSINFOpdf
\else
\fi
\usepackage{amssymb}
\usepackage{graphicx}
\usepackage{graphics}
\usepackage{color}
\usepackage{algorithm}
\usepackage{algorithmic}
\usepackage{amsmath}
\usepackage{cite}
\usepackage{soul}
\usepackage{cancel,comment}
\usepackage[normalem]{ulem}

\newcommand{\hs}[1]{\textcolor{magenta}{#1}}

\allowdisplaybreaks[3]

\begin{document}
%
\title{Mode Consensus Algorithms With Finite Convergence Time}
%
%
%

\author{Chao~Huang, Hyungbo~Shim, Siliang~Yu, Brian~D.~O.~Anderson \quad
\thanks{This work was supported in part by the National Natural Science Foundation of China under Grant 62373282, Grant 62350003 and Grant 62150026 and by the National Research Foundation of Korea grant funded by MSIT (No.~RS-2022-00165417).}
\thanks{Chao Huang and Siliang Yu are with the College of Electronic and Information Engineering, Tongji University, Shanghai 200092, China (e-mail: csehuangchao@tongji.edu.cn.)}
\thanks{Hyungbo Shim is with ASRI, Department of Electrical and Computer Engineering, Seoul National University, Seoul, Korea (e-mail: hshim@snu.ac.kr)}
\thanks{Brian D.O. Anderson is with the School of Engineering, Australian National University, Acton, ACT 2601, Australia. (e-mail: brian.anderson@anu.edu.au.)}
\thanks{Manuscript received XX XX, 20XX; revised XX XX, 20XX.}}

%
%

\markboth{Journal of \LaTeX\ Class Files,~Vol.~6, No.~1, January~2007}%
{Shell \MakeLowercase{\textit{et al.}}: Bare Demo of IEEEtran.cls
for Journals}
%



\maketitle

\begin{abstract}
This paper studies the distributed mode consensus problem in a multi-agent system, in which the agents each possess a certain attribute and they aim to agree upon the mode (the most frequent attribute owned by the agents) via distributed computation. Three algorithms are proposed. The first one directly calculates the frequency of all attributes at every agent, with protocols based on blended dynamics, and then returns the most frequent attribute as the mode. Assuming knowledge at each agent of a lower bound of the mode frequency as \emph{a priori} information, the second algorithm is able to reduce the number of frequencies to be computed at every agent if the lower bound is large. The third algorithm further eliminates the need for this information by introducing an adaptive updating mechanism. The algorithms find the mode in finite time, and estimates of convergence time are provided. The proposed first and second algorithms enjoy the plug-and-play property with a dwell time.
\end{abstract}

\begin{IEEEkeywords}
Consensus, Mode computing, Blended dynamics, Plug-and-play
\end{IEEEkeywords}

\newtheorem{Def}{Definition}
\newtheorem{Asu}{Assumption}
\newtheorem{Thm}{Theorem}
\newtheorem{Cor}{Corollary}
\newtheorem{Alg}{Algorithm}
\newtheorem{Rem}{Remark}
\newtheorem{Lem}{Lemma}
\newtheorem{Pro}{Problem}
\newtheorem{pot}{Proof of Theorem}

%
\IEEEpeerreviewmaketitle

\section{Introduction}

Distributed mode consensus, also known as majority voting or multiple voting, allows for the identification of the most frequent choice when dealing with categorical data like movies, car brands, or political candidates. Since it is not possible to directly calculate average or median values for such inherently nonnumerical data, distributed mode consensus provides a way to determine the central tendency. In the existing literature, achieving consensus on functions of interest, known as the $f$-consensus problem \cite{cortes2008distributed,huang2023distributed}, has been successful for specific types of functions typically assuming real values such as finding average, max(min), median, or the $k$-smallest element. While distributed convex optimization based protocols can handle consensus on these functions directly, the mode consensus problem seems an exception. In addition, the mode function cannot be represented as a composition of the functions mentioned above, presenting a non-trivial challenge for mode consensus.

Achieving mode consensus is not an entirely new problem of course. In the literature, Ref.~\cite{kuhn2008distributed} introduces a distributed method for computing the mode. In this method, the frequency of each element is aggregated from the leaves to the root along a spanning tree, and only the root node performs the mode calculation. By incorporating hash functions, the algorithm is able to find the mode with high probability and low time complexity. The binary majority voting problem, where there are only two distinct elements in the population, is addressed using the ``interval consensus gossip" approach described in \cite{benezit2009interval}. The state space used for this problem is $\{0, 0.5^-, 0.5^+, 1\}$. Initially, nodes vote for either ``$0$" or ``$1$" with corresponding states of $0$ or $1$. When neighboring nodes come into contact, they exchange their states and update them based on a predefined transition rule. When the algorithm reaches convergence, all nodes are expected to have states within the set $\{0, 0.5^-\}$ if ``$0$" is the majority choice. Conversely, if ``$1$" is the majority choice, all node states will belong to the set $\{0.5^+, 1\}$. Subsequently in \cite{benezit2011distributed}, a Pairwise Asynchronous Graph Automata (PAGA) has been used to extend the above idea to the multiple choice voting problem, and sufficient conditions for convergence are stated. In \cite{salehkaleybar2015distributed}, a distributed algorithm for multi-choice voting/ranking is proposed. The interaction between a pair of agents is based solely on intersection and union operations. The optimality of the number of states per node is proven for the ranking problem. Ref.~\cite{dashti2022distributed} explores distributed mode consensus in an open multi-agent system. Each agent utilizes an average consensus protocol to estimate the frequency of each element and then selects the one with the highest frequency as the mode. Agents are free to join or leave the network, and the mode may vary during the process, but the agent that leaves the network needs to signal this intention to its neighbors beforehand.

In this paper, we present distributed mode consensus algorithms based on the concept of blended dynamics \cite{lee2018distributed,lee2021design}. Blended dynamics have the characteristic that the collective behavior of multi-agent systems can be constructed from the individual vector fields of each agent when there is strong coupling among them. As an example, \cite{lee2018distributed} has demonstrated that individual agents can estimate the number of agents in the network in a distributed manner. The proposed mode consensus algorithms provide two main key benefits, over and beyond the inherent contribution. First, the algorithms can be easily implemented in a plug-and-play manner. This means that the system can maintain its mode consensus task without requiring a reset of all agents whenever a new agent joins or leaves the network. Second, we can demonstrate the intuitively satisfying conclusion that the frequency of the mode has an impact on the convergence rate of the mode consensus algorithm, in the sense that a higher mode frequency results in faster convergence of the algorithm.

The paper is organized as follows. In Section~\ref{sec:prel}, preliminaries on $f$-consensus is introduced, and the mode consensus problem is then described. The direct mode consensus algorithm is described in Section~\ref{sec:direct}, along with its characterization of convergence rate. Section~\ref{sec:main} combines the direct algorithm with the $k$-th smallest element consensus, resulting in two mode consensus algorithms that are applicable when the mode appears frequently. The performance of the proposed algorithms is evaluated in Section~\ref{sec:simul}. Finally, Section~\ref{sec:conclusion} concludes the paper.

\section{Notation and Preliminaries}\label{sec:prel}


\subsection{Underlying network}

Consider a group of $N$ agents labeled as $\mathcal V = \{1,\cdots,N\}$. 
Every agent has an \textit{attribute}, which can be thought of as a label. The attribute could be a positive integer, a real vector, a color, a gender, an age group, a brand of car, etc. Two or more agents may have the same attribute (and indeed, in many situations one might expect the number of distinct attributes to be much less than the number of agents). The attribute of agent~$i$ will be denoted by $a_i$. The vertex set is part of an undirected graph $\mathcal G$ with which is also associated a set $\mathcal E$ of edges (i.e. vertex pairs), in the usual way. The neighbor set $\mathcal N_i$ of agent $i$ is the set of vertices which share an edge with agent $i$.
The vertex set $\mathcal V$, the edge set $\mathcal E$, the attributes $a_i$, and $\mathcal N_i$ are assumed to be time-invariant in the bulk of the paper, but at times we open up the possibility, to accommodate a ``plug and play" capability, that, following certain rules, means they are piecewise constant.
The state $x_i$ is updated by an {\em out-distributed} control algorithm, that is, at every time $t$, the quantity $\dot x_i$ is computed as some function of $a_i$, $x_i(t)$, and $x_j(t)$ for all $j \in \mathcal N_i\left(t\right)$.\footnote{While we choose a continuous time setting in this paper, it seems very probable that a discrete-time setting could be used as an alternative, with very little change to the main assumptions, arguments and conclusions.}

\begin{Asu}\label{Asu:connectivity}
The graph $\mathcal G=(\mathcal V,\mathcal E)$ is undirected and connected, with $|\mathcal V|=N$. 
\end{Asu}

The need for connectivity is intuitively obvious. The need for $\mathcal G$ to be undirected is less so; note though that virtually all blended dynamics developments rest on an assumption of undirectedness of the underlying graph.

\subsection{$f$-consensus and broad problem setting}

To explain the problem setting, we define first the particular generalization of consensus, viz. $f$-consensus, with which we are working, and then indicate an explicit type of $f$-consensus problem of interest in this paper for which we are seeking an update algorithm. Following this, we provide two illustrative examples of $f$-consensus which are in some way relevant to the problem considered here.  For an introduction and the development of $f$-consensus, see e.g. \cite{olfati-Saber2007consensus,cortes2008distributed,huang2023distributed}.

\begin{Def}[$f$-consensus]\label{f-consensus}
With $\Omega$ being the set of all possible distinct attributes, consider a collection of $N$ agents whose attributes take values in $\Omega$.  
Suppose $f: \Omega^N \to \Omega$ is a given function. 
An algorithm is said to {\em achieve $f$-consensus asymptotically} if it is out-distributed and 
the solution of the overall system exists and satisfies, for every $i \in \mathcal V$, $\lim_{t \to \infty } x_i(t) = f(a_1,\cdots,a_N)$.
\end{Def}

Average consensus, obtained by setting $\Omega=\mathbb R$, $a_i=x_i(0)$ and $f(a_1,\cdots,a_N) = \sum_{i=1}^N a_i/N$ is a very well known example. 
Some others are detailed further below, starting with the problem of interest in this paper. 
In the meantime however, 
we shall make the following assumption:

\begin{Asu}\label{Asu:finiteattributes}
The set $\Omega$ is finite, and there is a bijective mapping $l: \Omega \to \mathcal{D} := \{ 1, 2, \dots, |\Omega|\}$.
\end{Asu}

To illustrate the practical effect of this assumption, suppose that we are considering an attribute which is a 2-vector of real numbers, being the height and weight of a group of individuals. 
Such data is always quantized in the real world, with height for example usually being expressed to the nearest number of centimeters. 
So the vector entry corresponding to height might be an integer number somewhere between 25 and 250, and a number of say 180 would indicate the individual's height lies in the interval [179.5,180.5). 
In effect $\Omega$ becomes a finite subset of $\mathbb R^2$.  
Then any such finite set $\Omega$ can always be bijectively mapped to $\{1,2,\cdots,|\Omega|\} = \mathcal{D}$.
While the possibly unordered set $\Omega$ is mapped to an ordered set $\mathcal{D}$ by the mapping $l$, the order can be arbitrary for our purpose of computing the mode of the attributes.

\subsection{Two relevant examples of $f$-consensus}

There are two related problems treatable by $f$-consensus which are similar to the problem just posed, and which have provided to the authors of this paper insights in the formulation of the solution of the mode consensus problem.

\subsubsection{Distributed computation of network size}

In the literature, there exist consensus protocols that accomplish the task of distributed computation of the network size based on blended dynamics, see e.g.~\cite{lee2018distributed}. 
Inspired by \cite{lee2018distributed}, the following simple protocol estimates $N$ in finite time under the assumption that $N \le \bar N$ where $\bar N$ is a known upper bound of $N$:
\begin{align}\label{Nconsensus}
\begin{split}
\dot x_1 &=h_x\left[ -x_1 + 1 + \gamma_x \sum_{j \in {\mathcal N}_1} (x_j - x_1)\right], \\
\dot x_i &=h_x\left[ 1 + \gamma_x \sum_{j \in {\mathcal N}_i} (x_j - x_i)\right] , \quad \forall i \not = 1,
\end{split}
\end{align}
where $\gamma_x>0$ is the coupling gain, and $h_x>0$ is the gain  to control the speed of the algorithm. 
As can be shown in Theorem \ref{thm1} of the next section, if $x_i(0) \in {\mathcal K}_x := [0.5, \bar N + 0.5]$ and $\gamma_x \ge \bar N^3$, the solution of the system \eqref{Nconsensus} satisfies
\[\langle x_i(t) \rangle = N, \quad \forall t > {\mathcal T}_x\]
where $\left\langle \cdot \right\rangle$ is the rounding function, and
$${\mathcal T}_x = \frac{4 \bar N}{h_x} \ln \frac{4  M_{{\mathcal K}_x} \sqrt{\bar N}}{2-\sqrt{2}}$$
where $M_{{\mathcal K}_y} = \bar N$.


\begin{Rem}
The upper bound on the estimation time ${\mathcal T}_x$ (which may be quite conservative) depends on $\bar N$ in the order of $O(\bar N \ln \bar N)$.
If $h_x$ grows linearly with $\bar N$, one may even have $O\left(\ln {\bar N}\right)$. However a large gain could also undermine the robustness of the protocol against high-frequency noise.
\end{Rem}

\subsubsection{$k$-th smallest element consensus}

Since $\mathcal D$ is a totally ordered set, suppose without loss of generality that $l(a_1) \le l(a_2) \le \cdots \le l(a_N)$, then the $k$-th smallest element is defined as $a_k$. 
The $k$-th smallest element (or $k$-th order statistic) consensus problem is then an $f$-consensus problem with $f(a_1,\cdots,a_N) = a_k$.

Ref.~\cite{huang2023distributed} proposed a method to solve the $k$-th smallest element consensus problem with distributed convex optimization algorithms. 
An example used in the following is 
\begin{equation}\label{k-th1}
\dot z_i = -\phi_k(z_i,a_i,N) + \gamma_z \sum \limits_{j \in {\mathcal N_i}} {\rm{sgn}}\left( z_j - z_i \right) ,
\end{equation}
where $\phi_k : {\mathbb R} \times \Omega \times {\mathbb N}$ is defined by
\begin{equation*}
\phi_k(z_i,a_i,N) = \begin{cases}
\beta (z_i - l(a_i)) - gk, & z_i < l(a_i), \\
0, & z_i = l(a_i), \\
\beta (z_i - l(a_i)) + g \left( N + 1 - k \right), & z_i > l(a_i),
\end{cases}
\end{equation*}
with $\beta > 0$ and $g > \beta \bar N \left|\Omega\right|$.\footnote{In \cite{huang2023distributed} the bound was given by $g > \beta N \left| {{l(a_k)} - \frac{1}{N}\sum\nolimits_{i = 1}^N {{l(a_i)}} } \right|$, which can be loosened to $g > \beta N \left| {{l(a_N)} - {{l(a_1)}} } \right|=\beta N\left|\Omega\right|$.}\footnote{System \eqref{k-th1} admits unique solution in the sense of Filippov. For more details, refer to, e.g., \cite[Proposition S2]{cortes2008discontinuous}.
}
Initial conditions for the differential equations can be arbitrary. It can be shown following \cite{Li2022exponentially} that if 
\begin{equation}\label{k-th3}  
\gamma_z > \bar N {\max _{1\le i\le N}}{\sup_{\tau \in {\mathcal K}_z}} \left| \phi_k(\tau, a_i, N) \right|,
\end{equation}
the convergence rate of the protocol 
\eqref{k-th1}--\eqref{k-th3} satisfies $V(t) \le e^{ - \beta t} V(0)$, where $V(t) = \frac{1}{2} \sum_{i = 1}^N \left|z_i(t) - l(a_k)\right|^2$. When $\mathcal K_z=\left[ 0.5, \bar N + 0.5 \right]$, 
it can be obtained that the solution of \eqref{k-th1}--\eqref{k-th3} satisfies
\[\left\langle z_i(t) \right\rangle = l(a_k),\quad \forall t > {\mathcal T_z}\]
where
\[{\mathcal T_z} = \frac{1}{\beta }\ln {2 \bar N \left| \Omega \right|}.\]
By inverting the map $l$, every agent can figure out the $k$-th smallest element.

\begin{Rem}
Although increasing $\beta$ renders a smaller $\mathcal T_z$, it also amplifies $\gamma_z$ to a very large number for large $\bar N$. To reach a balance between $\mathcal T_z$ and $\gamma_z$, we select $\beta=O\left(\frac{1}{\bar N}\right)$, so that $\mathcal T_z=O\left(\bar N \ln \bar N\right)$ and $\gamma_z=O\left(\bar N^2\right)$.
\end{Rem}

\begin{Rem}
Both of the examples above, the distributed computation of network size and the $k$-th smallest element consensus, exhibit algorithms with a finite convergence time. Likewise, for mode consensus, we seek algorithms which also yield finite consensus time. 
\end{Rem}

\subsection{The problem of interest: Mode consensus} 

Mode consensus is a special class of $f$-consensus problem. 
Suppose the function $f_{\rm m}:\Omega^N \to \Omega$ returns the attribute $a^* \in \Omega$ that appear most often among $a_1,\cdots,a_N$ (when multiple distinct values equally appear most often, $f_{\rm m}$ returns any of these values specified by the user). 
The mode consensus problem as studied in this paper is an $f$-consensus problem with $f=f_{\rm m}$.

\subsection{Plug-and-play}\label{sec:pnp}

In many circumstances, it may be advantageous to have a plug-and-play capability. Specifically, 
we are interested in a network that can change over time during its operation.
Changes cannot be arbitrary, but rather admissible in accord with certain rules.  

First, while the \textit{potential} vertex set $\bar{\mathcal V}$ is taken as time-invariant, that is, $\bar{\mathcal V} = \{1, \cdots, \bar N \}$ is fixed over time for some $\bar N$, the actual vertex set $\mathcal V$ which is an arbitrary subset of $\bar{\mathcal V}$ can be time-varying but piecewise constant; with $N(t)$ denoting its cardinality, $\bar N$ is an upper bound for $N(t)$.
In this setting, our {\em admissible changes} are as follows:

\begin{itemize}
	\item The set of edges, written as $\mathcal E(t)$, is time-varying but piecewise constant.  Therefore, at certain time instants, a new edge or edges can be created, and some existing edge or edges can be deleted. 
	\item There can be orphan nodes (meaning a node that has no incident edge). When all the edges that are incident to a node are deleted, we say the node leaves the network.
	\item We assume that there is only one connected component in the network. In fact, even if there is more than one connected component, our concern is with just one of them. If a connected component of interest splits into two connected components (with some edges being deleted), one of these components still receives our attention, and all the nodes belonging to the other component are regarded as leaving the network.
	\item The attribute of a node is permitted to be  time-varying but must be piecewise constant.
	\item The node dynamics stop integrating whenever the node is an orphan. One example of the node dynamics is:
    \begin{equation}\label{eq:ppexam}
	\dot x_i = {\mathrm{sgn}}(|\mathcal N_i(t)|) \cdot \left( g_i(x_i,t) + \sum_{j \in \mathcal N_i(t)} (x_j - x_i) \right) 
    \end{equation}
	where $|\cdot|$ for a set implies cardinality of the set.
	In this way, malfunctioning agents can also be represented by considering them as orphans.
\end{itemize}

The control algorithm for updating the states of agents is said to be \textit{plug-and-play ready} if, whenever an abrupt admissible change of the network occurs, (a) the $f$-consensus is recovered (to a new value reflecting the new situation) after a transient, and (b) the recovery is achieved by {\em passive manipulation}, together with {\em local initialization} of the newly joining agent if necessary. By {\em passive manipulation} we mean that, when an individual agent detects the changes in its incident edge set, or equivalently its neighbor set, associated with immediate neighbors leaving or joining the network, the agent can perform some actions that do not require reactions of neighboring agents. An example is \eqref{eq:ppexam} because, if $\mathcal N_i(t^-) \not = \mathcal N_i(t^+)$ at time $t$, the manipulation simply resets the incident edge set or equivalently neighbor set, and this can be done by the individual agent. (Of course, neighbor agents may have to also reset in order to carry out their own updates.) By {\em local initialization} we mean that, any initialization following a change must be local only, i.e. it must be {\em global initialization-free}.  This implies that the algorithm should not depend on a particular initial condition constraining two or more agents in some linked way.
A constraint such as $\sum_{i=1}^N x_i(0) = 0$ is an example of global initialization, while the constraint that $x_i(0) \in {\mathcal K}$, $\forall i \in {\mathcal V}$, where $\mathcal K$ is a compact set known to every agent, is considered as an example of local initialization (or equivalently it is global initialization-free), and a local initialization is required for the newly joining agent (or, in \eqref{eq:ppexam}, when $\mathcal N_i(t)$ becomes non-empty so that the sign function changes from 0 to 1).

The property of plug-and-play ready basically requires that, when the change occurs, no further action is required except for the agents whose incident edge sets are changed.
In particular,  the requirement of passive manipulation is useful in the case when some agent suddenly stops working (without any prior notification) and cannot function properly anymore.

\section{Direct mode consensus algorithm}\label{sec:direct}

The following protocol, which is motivated by the algorithm for distributed computation of network size discussed above, lays the foundation of the mode consensus algorithm.  
It is also inspired by the notion of blended dynamics and calculates the number of agents with an arbitrary particular attribute $a \in \Omega$, denoted by $\mathcal F(a)$, in a distributed manner:
\begin{align}\label{Mode-consensus}
\begin{split}
\dot y_1 &= h_y\left[-y_1 + I(a,a_1) + \gamma_y \sum_{j \in {\mathcal N}_1} (y_j - y_1)\right], \\
\dot y_i &= h_y\left[I(a,a_i) + \gamma_y \sum_{j \in {\mathcal N}_i} (y_j - y_i)\right], \qquad \forall i \not = 1
\end{split}
\end{align}
where
\begin{equation}\label{ell}
I(a,a_i) = \begin{cases} 1, & a_i = a, \\ 0, & a_i \ne a, \end{cases}
\end{equation}
and analogous to (\ref{Nconsensus}), here $\gamma_y>0$ is the coupling gain, and $h_y>0$ is the gain to control the speed of the algorithm.

\begin{Thm}\label{thm1}
Suppose Assumptions \ref{Asu:connectivity} and \ref{Asu:finiteattributes} hold. If $\gamma_y \ge \bar N^3$, then for any initial condition $y_i(0) \in {\mathcal K}_y := [-0.5,\bar N+0.5]$, the solution of the consensus protocol \eqref{Mode-consensus}--\eqref{ell} satisfies
\begin{equation}\label{T3}
\left\langle y_i(t) \right\rangle = \mathcal F(a), \quad \forall t > {\mathcal T}_y,
\end{equation}
where, with $M_{{\mathcal K}_y} = \bar N + 1$ (the size of ${\mathcal K}_y$),
$${\mathcal T}_y = \frac{4 \bar N}{h_y} \ln \frac{4 M_{{\mathcal K}_y} \sqrt{\bar N}}{2-\sqrt{2}}.$$
\end{Thm}

Proof of Theorem \ref{thm1} is found in the Appendix.

\begin{Rem}
If $a=a_1=\cdots=a_N$, there is no essential difference between (\ref{Nconsensus}) and (\ref{Mode-consensus}). Thus the proof of Theorem \ref{thm1} is also applicable to the network size estimation protocol (\ref{Nconsensus}). 
\end{Rem}

\begin{Rem}
In the literature, $\mathcal F(a)/N$ can be estimated using average consensus protocols, see \cite{dashti2022distributed}. However, in order to run the protocol in an open multi-agent system, the agent that leaves the network must signal its intention to its neighbors beforehand. This is not required by the proposed Algorithm~\ref{Alg1}.
\end{Rem}

\begin{Rem}
As is usual for algorithms based on using the blended dynamics approach, grounded in singular perturbation ideas, there are broadly speaking two convergence rates, the fast one being associated with the achieving of consensus between the agents (adjusted by $\gamma_y$), and the slower one being associated with the blended dynamics (adjusted by $h_y$). This behavior is apparent in the simulations discussed later.
\end{Rem}

\begin{algorithm}
	\caption{Distributed mode consensus algorithm run at every agent $i$}
	\label{Alg1}

 \begin{enumerate}
     
     \item Run the distributed consensus protocol \eqref{Mode-consensus}--\eqref{ell} with $y_i(0) \in {\mathcal K}_y$ to estimate ${\mathcal F}(a)$ for every $a \in \Omega$;

     \item Return the attribute defined by the mode:
     \begin{equation}\label{eq:al1}
     a^* = \arg\max_{a \in \Omega} {\mathcal F}(a) .
     \end{equation}

 \end{enumerate}
 \end{algorithm}

Based on Theorem \ref{thm1}, Algorithm~\ref{Alg1} for distributed computation of the mode can be formulated directly. Evidently, one can execute the second step of Algorithm~\ref{Alg1} using parallel computations, with one for each attribute.  
Consider the following modification to \eqref{Mode-consensus}--\eqref{ell}:
\begin{align}\label{eq:counter}
\begin{split}
    \dot \xi_1 &= h_\xi\left[-\xi_1 + e_{l(a_1)} + \gamma_\xi \sum_{j \in \mathcal N_1} (\xi_j-\xi_1)\right] \\
    \dot \xi_i &= h_\xi\left[e_{l(a_i)} + \gamma_\xi \sum_{j \in \mathcal N_i} (\xi_j-\xi_i)\right], \qquad \forall i = 2,\cdots,|\Omega|,
\end{split}
\end{align}
where $\xi_i \in \mathbb R^{|\Omega|}$ is the state vector, and $e_i \in \mathbb R^{|\Omega|}$ is the unit $|\Omega|$-vector. 
Likewise, $\gamma_\xi$ and $h_\xi$ are positive scalar gains. 
If each agent holds the set $\Omega$ and the map $l$, then the execution of the third step of the algorithm is straightforward (and achievable by a single agent, or all agents).
Any agent $i$ for which $a_i=a^*$ by definition has an attribute which is the mode.

\begin{Rem}
Indeed, if there is more than one attribute with the highest frequency of occurrence, Algorithm \ref{Alg1} is able to find all of these attributes. In that case the user has the option to return any one or several of them.
This issue will not be examined any further in the remainder of the paper.   
\end{Rem}

Algorithm \ref{Alg1} is attractive on several grounds. 
It offers finite convergence time with a bound available for that time, and it is plug-and-play ready.
In particular, if the admissible changes discussed in Section \ref{sec:pnp} occur with the dwell time ${\mathcal T}_y$ (i.e., any two consecutive changes do not occur within ${\mathcal T}_y$), then the mode is obtained after the time ${\mathcal T}_y$ from the time of change.
It is because, whenever $y_i(t_j) \in {\mathcal K}_y$ where $t_j$ is the $j$-th time of change, it holds that $-0.5+{\mathcal F}(a) < y_i(t) < {\mathcal F}(a)+0.5$ for all $t \ge t_j + {\mathcal T}_y$ by Theorem \ref{thm1}.
Thus, $y_i(t_{j+1}) \in {\mathcal K}_y$ because ${\mathcal F}(a) \in [1,\bar N]$, and this repeats.
On the other hand, it has the potential disadvantage that every agent has to run the consensus protocols \eqref{Mode-consensus}--\eqref{ell} multiple (in fact $|\Omega|)$ times, which could be computationally burdensome (but may not be, even with large $N$).
For small $|\Omega|$, there is in fact no substantive disadvantage.


\section{Mode consensus algorithm considering the frequency of mode}\label{sec:main}

In this section, we consider two alternative algorithms based on knowledge, available \textit{a priori} or acquired early in the algorithm, of a lower bound on the frequency of the mode.

This second style of mode consensus algorithm in both cases uses the following result. 

\begin{Lem}\label{lem1}
Let $a \in \Omega$, and let $K$ be a positive integer. 
If $\mathcal F(a) \ge \left\lceil \frac{N}{K} \right\rceil$, then there is an integer $j \in \left\{1,2,\cdots,K\right\}$ such that $l(a)$ equals the $j\left\lceil \frac{N}{K} \right\rceil$-th smallest element of $\{l(a_1),\cdots,l(a_N)\}$.
\end{Lem}

\begin{IEEEproof}
The proof uses a typical pigeonhole principle argument. 
For notational convenience, let $l_i := l(a_i)$ and $l_a := l(a)$.
Suppose without loss of generality that $l_1 \le \cdots \le l_N$. 
Moreover, let $l_{N+1}, \cdots, l_{K\left\lceil \frac{N}{K} \right\rceil}$ be additional attributes introduced temporarily just for the proof and such that 
\begin{equation}\label{order}
l_1 \le \cdots \le l_N \le l_{N+1} \le \cdots \le l_{K\left\lceil \frac{N}{K} \right\rceil}.
\end{equation}
Partition the above sequence into subsequences 
$${\mathcal D}_j = \left\{l_{\left(j-1\right)\left\lceil {\frac{N}{K}} \right\rceil+1}, l_{\left(j-1\right)\left\lceil {\frac{N}{K}} \right\rceil+2}, \cdots, l_{j\left\lceil {\frac{N}{K}} \right\rceil}\right\}$$ 
for $j=1,2,\cdots,K$. 
It follows that $\left|{\mathcal D}_j\right| = \left\lceil {\frac{N}{K}}\right\rceil$.

The result is then proved with a contradiction argument. 
Suppose, to obtain a contradiction, that $l_a \neq l_{j\left\lceil {\frac{N}{K}} \right\rceil}$ for all $j \in \left\{1,2,\cdots,K\right\}$. 
Then it must follow that $l_a = l_{\left(j-1\right)\left\lceil {\frac{N}{K}} \right\rceil + s}$ for some $j \in \left\{1,2,\cdots,K\right\}$ and $s \in \left\{1,\cdots,\left\lceil {\frac{N}{K}}\right\rceil-1\right\}$. 
Combined with the fact that the values are ordered (see \eqref{order}), it must follow that the number of agents with attribute equal to $a$ is less than $\left\lceil {\frac{N}{K}}\right\rceil$, which yields a contradiction. 
\end{IEEEproof}

\begin{Rem}\label{rem:3}
In Lemma \ref{lem1}, if in addition, we have $\left\lceil {\frac{N}{K}} \right\rceil > {\frac{N}{K}} $, the result of Lemma \ref{lem1} can be mildly strengthened to requiring $j\in\left\{1,2,\cdots,K-1\right\}$, because the $j\left\lceil {\frac{N}{K}} \right\rceil$-th smallest element with $j=K$ is then out of the set $\{l(a_1),\cdots,l(a_N)\}$.
\end{Rem}

\subsection{Algorithm with a priori knowledge of the least frequency of the mode}

The message of Lemma~\ref{lem1} is that, if a lower bound of the frequency of the mode is known, that is, ${\mathcal F}(a^*) \ge f^*$ with a known $f^*$, then, with $K$ such that $f^* \ge \lceil \frac{N}{K} \rceil$, the integer $l(a^*)$ for the mode $a^*$ should be one of the $j\left\lceil {\frac{N}{K}} \right\rceil$-th smallest elements, $j=1,\dots,K$, in $\{l(a_1),\dots,l(a_N)\}$.
This message yields Algorithm~\ref{Alg2}, in which, Step 2) identifies the $j\left\lceil {\frac{N}{K}} \right\rceil$-th smallest elements, $j=1,\dots,K$, and then, Step 4) finds the mode by comparing the frequencies among the candidates.

\begin{algorithm}
	\caption{Distributed mode consensus algorithm run at every agent $i$}
	\label{Alg2}

\begin{enumerate}
     
     \item Estimate the network size, $N$, with the distributed consensus protocol \eqref{Nconsensus} with $x_i(0) \in {\mathcal K}_x$;
     
      \item For each $j\in\left\{1,2,\cdots,K\right\}$ (or $j\in\left\{1,2,\cdots,K-1\right\}$ if $\left\lceil {\frac{N}{K}} \right\rceil > {\frac{N}{K}} $), run the consensus protocol \eqref{k-th1}--\eqref{k-th3} with $z_i(0) \in {\mathcal K}_z$ to estimate the $j\left\lceil {\frac{N}{K}} \right\rceil$-th smallest element $\alpha_j \in \Omega$.      
      Collect them as ${\mathcal A} := \{ \alpha_1, \dots, \alpha_{|{\mathcal A}|} \}$ where $|\mathcal A| \le K$ or $K-1$.
     
     \item Run the distributed consensus protocol \eqref{Mode-consensus}--\eqref{ell} with $y_i(0) \in {\mathcal K}_y$ to estimate ${\mathcal F}(\alpha)$ for every $\alpha \in {\mathcal A}$;
     
     \item  Return the mode
     \begin{equation}\label{eq:al2}
     a^* = \arg\max_{\alpha \in {\mathcal A}} \left\{ {\mathcal F}(\alpha) \right\}.
     \end{equation}
\end{enumerate}
\end{algorithm}

While Algorithm \ref{Alg2} outlines the distributed mode consensus algorithm, it also reveals the significance of $K$ in reducing the computational load.
Depending on $K$, sometimes Algorithm~\ref{Alg2} may involve manipulating or storing fewer variables than Algorithm~\ref{Alg1}, but the reverse may hold. 
In fact, Step 2) in Algorithm~\ref{Alg2} can be considered as a selection procedure to find an attribute to be inspected.
By this, it is expected that the number of inspections in \eqref{eq:al2} (effectively, the cardinality of $\mathcal A$) is smaller than that in \eqref{eq:al1} (the cardinality of $\Omega$). This is indeed the case when, for example, with $N=10$, $[l(a_1),\dots,l(a_N)] = [1,1,1,1,1,2,3,4,5,6]$, and $|\Omega| = 6$.
That is, for \eqref{eq:al1}, at least six attributes are inspected in Step 1) of Algorithm~\ref{Alg1}.
For the case of Algorithm~\ref{Alg2}, assuming that $f^*=5$ is known, $K=2$ guarantees that $f^* = 5 \ge \lceil 10/2 \rceil = 5$, and so two attributes are inspected by Step 3) of Algorithm~\ref{Alg2}.
However, in order to identify the attributes to be inspected, Step 1)-2) of Algorithm~\ref{Alg2} needs to be performed, as an overhead. So the total number of variables to be manipulated is $2K+1=5$ which is still less than $\left|\Omega\right|=6$.
As a second example, if $[l(a_1),\dots,l(a_N)] = [1,2,3,4,4,5,6,7,8,8]$ with $N=10$ and $\left|\Omega\right|=8$, then $f^*=2$ and we have to choose $K=5$.
In this case, Algorithm~\ref{Alg2} inspects five candidates while Algorithm~\ref{Alg1} inspects eight candidates, but the count of the overhead is five so that the total count becomes $2K+1=11$, which is more than $|\Omega|=8$.

A convergence time bound can be established as follows. Suppose Algorithm \ref{Alg2} is executed with parallel computations at Steps 2) and 4), for example at Step 2), each agent runs the $j\left\lceil {\frac{N}{K}} \right\rceil$-th smallest element consensus protocol for every $j=1,2,\cdots,K$ in parallel. Then, the mode consensus is reached within the time ${\mathcal T}_x + {\mathcal T}_y + {\mathcal T}_z$.

It appears from Algorithm \ref{Alg2} that Step $i+1)$ can only be implemented after Step $i)$ has converged. 
This is actually not the case. 
All steps may start simultaneously, provided that the parameters used in any step are replaced with their estimated value generated in the previous steps.
Indeed, the following equations are one alternative implentation of Algorithm~\ref{Alg2} for agent $i$:
\begin{align}\label{eq:alternative}
\begin{split}
    \dot x_i &= h_x\left[c_i x_i + 1 + \gamma_x \sum_{j \in \mathcal{N}_i} (x_j - x_i)\right] \\
    \dot z_1^i &= -\phi_1^K(z_1^i,a_i,x_i) + \gamma_z \sum_{j \in \mathcal{N}_i} {\mathrm{sgn}} (z_1^j - z_1^i) \\
    &\vdots \\
    \dot z_K^i &= -\phi_K^K(z_K^i,a_i,x_i) + \gamma_z \sum_{j \in \mathcal{N}_i} {\mathrm{sgn}} (z_K^j - z_K^i) \\
    \dot y_1^i &= h_y\left[c_i y_1^i + {\mathcal I}(z_1^i,a_i) + \gamma_y \sum_{j \in \mathcal{N}_i} (y_1^j - y_1^i)\right] \\
    &\vdots \\
    \dot y_K^i &= h_y\left[c_i y_K^i + {\mathcal I}(z_K^i,a_i) + \gamma_y \sum_{j \in \mathcal{N}_i} (y_K^j - y_K^i) \right]\\
    \widehat m^i &= \text{argmax}_{1 \le j \le K} \{ y_j^i \}
\end{split}
\end{align}
where $c_1=1$, $c_i=0$ for all $i \not = 1$, and 
\begin{align*}
    &\phi_k^K(z,a,x) \\
    &\qquad = \begin{cases} 
    \beta (z - l(a)) - g k \left\lceil \frac{\langle x \rangle}{K} \right\rceil, & z < l(a), \\
    0, & z = l(a), \\
    \beta (z - l(a)) + g \left( \langle x \rangle + 1 - k \left\lceil \frac{\langle x \rangle}{K} \right\rceil \right), & z > l(a), 
    \end{cases} \\
    &{\mathcal I}(z,a) = \begin{cases} 1, & \langle z \rangle = l(a) \\
    0, & \langle z \rangle \not = l(a) \end{cases}
\end{align*}
in which, $\beta$, $g$, $\gamma_x$, $\gamma_y$, and $\gamma_z$ are predetermined.
(When $\left\lceil {\frac{N}{K}} \right\rceil > {\frac{N}{K}} $, the $K$ in the above is interpreted as $K-1$.)
Here, $x_i$ is the estimate of $N$ by the agent $i$, $z_k^i$ is the estimate of the $k$-th smallest element, $y_k^i$ is the estimate of the frequency of $z_k^i$, and $\widehat m^i$ is the estimated integer for the mode, i.e., the estimated mode is $a \in \Omega$ such that $l(a) = \widehat m^i$. In fact, although all the dynamics \eqref{eq:alternative} run altogether, the estimates are sequentially obtained.
That is, $z_k^i$ starts converging to its proper value after $\langle x_i \rangle$ becomes $N$, and $y_k^i$ starts converging to its proper value after $\langle z_k^i \rangle$ becomes the $k$-th smallest element.
Therefore, the required time for getting the mode is still the same as ${\mathcal T}_x + {\mathcal T}_y + {\mathcal T}_z$.

Algorithm \ref{Alg2} that repeats forever, and the alternative algorithm \eqref{eq:alternative} are plug-and-play ready as long as the admissible changes occur with the dwell time ${\mathcal T}_x + {\mathcal T}_y + {\mathcal T}_z$.
This is because, with $x_i(t_j) \in {\mathcal K}_x$, $z_k^i(t_j) \in {\mathcal K}_z$, and $y_k^i(t_j) \in {\mathcal T}_y$ at the time of change $t_j$, the same inclusion holds at the next time of change $t_{j+1}$; i.e., $x_i(t_{j+1}) \in {\mathcal K}_x$, $z_k^i(t_{j+1}) \in {\mathcal K}_z$, and $y_k^i(t_{j+1}) \in {\mathcal T}_y$.



\vspace{3cm}

From (\ref{eq:alternative}) we see the number of the state variables needed in Algorithm \ref{Alg2} equals $2K+1$ (or $2K-1$ if $\left\lceil {\frac{N}{K}} \right\rceil > {\frac{N}{K}}$). Thus Algorithm \ref{Alg2} uses less state variables than Algorithm \ref{Alg1} if $2K+1<\left|\Omega\right|$ (or $2K-1<\left|\Omega\right|$ when $\left\lceil {\frac{N}{K}} \right\rceil > {\frac{N}{K}} $). In particular, If $K=1$, the mode consensus problem reduces to the max consensus of $\{l(a_1),\dots,l(a_N)\}$ since $l(a^*)$ is the $N$-th smallest element; If $K=2$, the problem reduces to finding the attributes having larger frequency between the median and the maximum of $\{l(a_1), \dots, l(a_N)\}$ (if $N$ is odd, it is simply the median); If $K=|\Omega|$, Algorithm~\ref{Alg2} probably has to do the $k$-th smallest element consensus $|\Omega|$ times, and then it offers no advantage over Algorithm~\ref{Alg1}. Since
$f^* \ge \left\lceil \frac{N}{|\Omega|} \right\rceil$, it is not necessary to consider $K > |\Omega|$ because the condition $f^* \ge \left\lceil \frac{N}{K} \right\rceil$ is automatically fulfilled. Thus to choose Algorithm~\ref{Alg2} over Algorithm~\ref{Alg1} it is preferable to have $2K+1 \ll |\Omega|$.

\subsection{Algorithm with learned knowledge of the least frequency of the mode}

When $f^*$ is unknown, we are not able to employ Algorithm~\ref{Alg2}.
However, once we determine a value of ${\mathcal F}(a)$ for some attribute $a$, then we can immediately infer that $f^* \ge {\mathcal F}(a)$.
Based on this simple fact, we begin with an estimate $F$ of $f^*$ with $F=1$, and update $F$ by ${\mathcal F}(a)$ whenever an attribute $a$ such that ${\mathcal F}(a) > F$ is found. 
At the same time, we begin with $K=1$ and repeatedly increase $K$ until it holds that $F \ge \lceil \frac{N}{K} \rceil$.
Once we have $F \ge \lceil \frac{N}{K} \rceil$, it means that the integer corresponding to the mode $a^*$ is among the $j\left\lceil {\frac{N}{K}} \right\rceil$-th smallest elements, $j=1,\dots,K$, of $\{ l(a_1),\dots,l(a_N)\}$, so that the previous algorithm can be applied.
Putting all this together, we obtain Algorithm~\ref{Alg3}.


\begin{algorithm}
\caption{Distributed mode consensus algorithm run at every agent $i$}
\label{Alg3}

 \begin{enumerate}
     
     \item Estimate the network size, $N$, with the distributed consensus protocol \eqref{Nconsensus} with $x_i(0) \in {\mathcal K}_x$. (If $N=1$, stop because the mode is the same as the attribute.);

    \item Set $K=F=1$;
     
    \item \textbf{While} $F < \left\lceil {\frac{N}{{{K}}}} \right\rceil $ \textbf{do} 

    \item $K \leftarrow K+1$;
     
    \item For each $j\in\left\{1,2,\cdots,K\right\}$ (or $j\in\left\{1,2,\cdots,K-1\right\}$ if $\left\lceil {\frac{N}{K}} \right\rceil > {\frac{N}{K}} $), 
    run the consensus protocol \eqref{k-th1}--\eqref{k-th3} with $z_i(0) \in {\mathcal K}_z$ to estimate the $j\left\lceil {\frac{N}{K}} \right\rceil$-th smallest element $\alpha_j \in \Omega$. Collect them as ${\mathcal A} := \{ \alpha_1, \dots, \alpha_{|{\mathcal A}|} \}$ where $|\mathcal A| \le K$ or $K-1$.

    \item Run the distributed consensus protocol \eqref{Mode-consensus}--\eqref{ell} with $y_i(0) \in {\mathcal K}_y$ to estimate ${\mathcal F}(\alpha)$ for every $\alpha \in {\mathcal A}$; 

    \item $F \leftarrow \max\left\{F, \max_{\alpha \in {\mathcal A}} \{ {\mathcal F}(\alpha) \} \right\}$; 

    \item \textbf{End While}   
     
    \item Return the mode
     \begin{equation*}
     a^* = \arg\max_{\alpha \in {\mathcal A}} \left\{ {\mathcal F}(\alpha) \right\}.
     \end{equation*}

\end{enumerate}
\end{algorithm}

In the algorithm, Steps 5) and 6) can be made more efficient by storing data obtained in the previous loop, but 
in the worst case scenario, the ``\textbf {While}" loop should be run $K^*$ times, where $K^*$ is the smallest positive integer such that $f^*\ge\lceil\frac{N}{K^*}\rceil$. Analogously to the analysis for Algorithm \ref{Alg2}, the number of state variables needed in Algorithm \ref{Alg3} equals $1+\sum\nolimits_{i = 1}^{{K^*}} 2i=K^*\left(K^*+1\right)+1$. To choose Algorithm~\ref{Alg3} over Algorithm~\ref{Alg1} it is preferable to have $K^*\left(K^*+1\right)+1 \ll |\Omega|$. If in addition, Algorithm \ref{Alg3} is executed with parallel computations, the time to reach mode consensus is less than ${\mathcal T}_x + K^*\left({\mathcal T}_y + {\mathcal T}_z \right)$.


\section{Simulations}\label{sec:simul}

Consider an undirected loop composed of $N=40$ agents, where agent $i$ is connected to agent $i+1$ for every $i=1,2,\cdots,N-1$, and connected to agent $i-1$ for every $i=2,3,\cdots,N$. Agent $1$ and $N$ are also  connected. Suppose that $\Omega = \{1,2,\cdots,10\}$, and
$$[\mathcal F(1), \mathcal F(2), \dots \mathcal F(10)] = [5, 6, 7, 16, 1,1,1,1,1,1].$$
Thus $a^*=4$ is the mode, which is unique.

\begin{figure}
\centering
\includegraphics[width=1.0\linewidth]{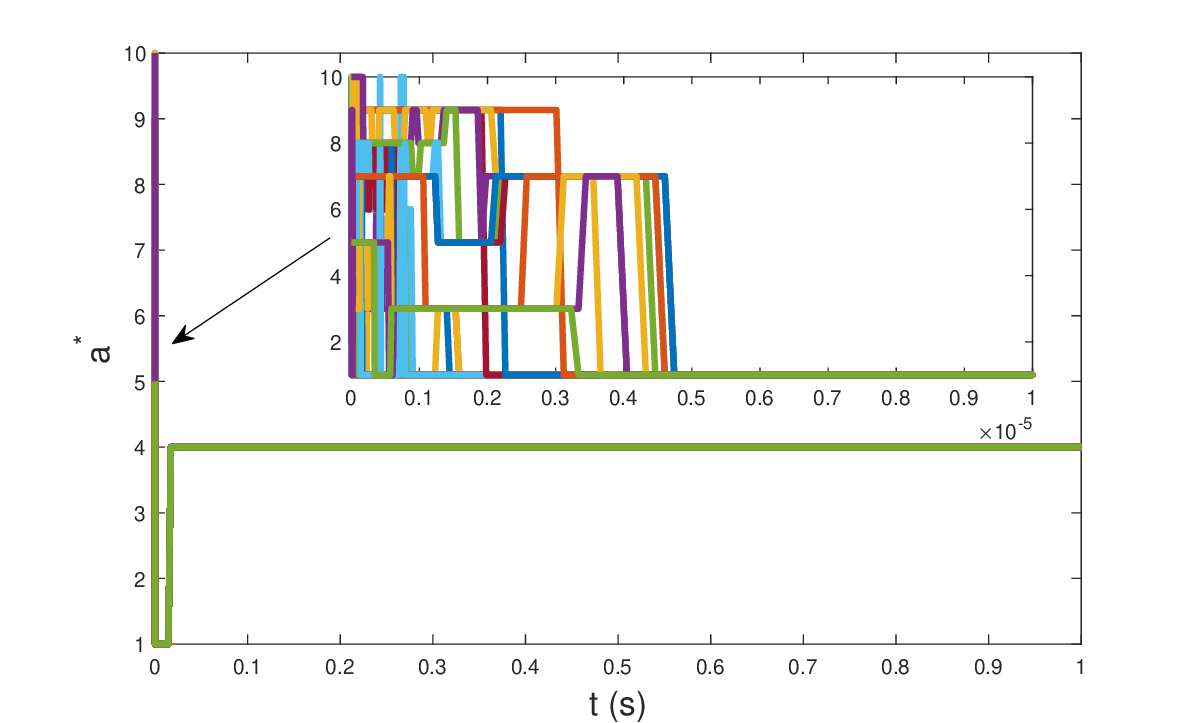}
\caption{The mode $a^*$ estimated at each agent with Algorithm \ref{Alg1}, converging to $4$.}
\label{Al1_1}
\end{figure}

\begin{figure}
\centering
\includegraphics[width=1.0\linewidth]{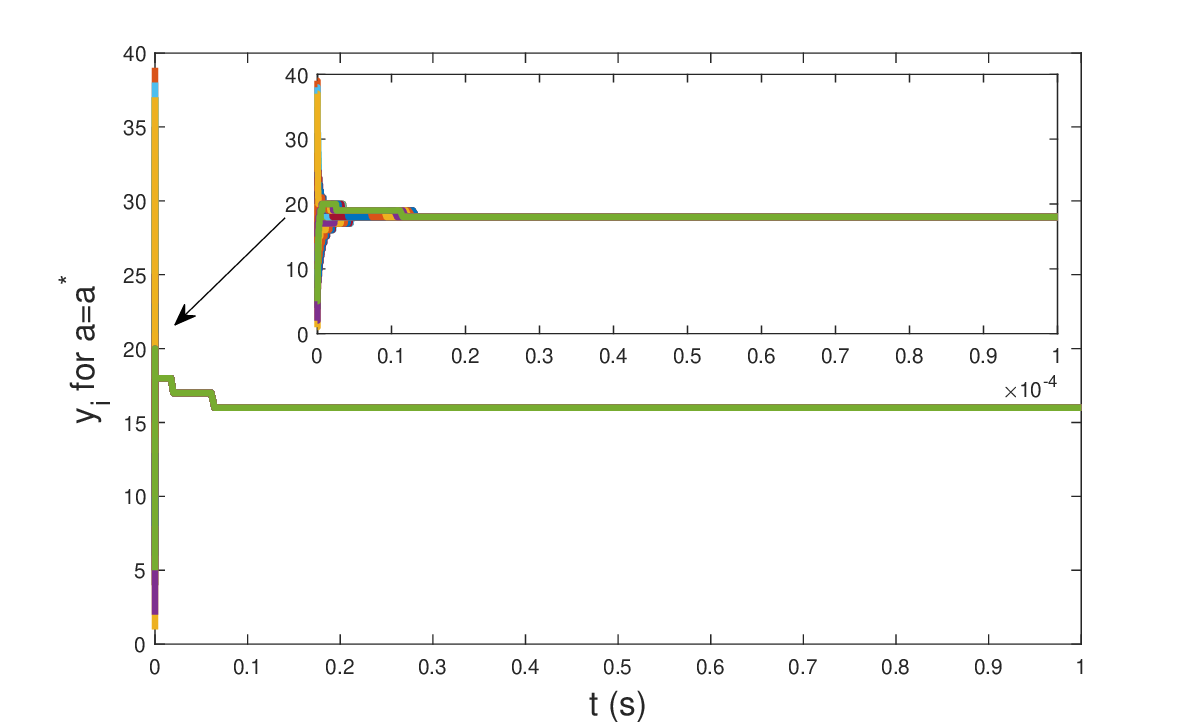}
\caption{The estimated frequency of the mode, i.e., $\mathcal F\left(a^*\right)$, at each agent with Algorithm \ref{Alg1}, converging to $16$.}
\label{Al1_2}
\end{figure}

Algorithm \ref{Alg1} is examined first. For each $a\in\Omega$, the initial condition of (\ref{Mode-consensus})--(\ref{ell}) at each agent (i.e., the initial guess of $\mathcal F\left(a\right)$ at each agent) is randomly and independently selected from $1$ to $40$.
According to Theorem \ref{thm1}, the coupling gain is selected as $\gamma_y = N^3= 6.4\times10^4$. Set $h_y = 10^3$.
The simulation results are shown in Figs.~\ref{Al1_1}--\ref{Al1_2}. 
From Fig.~\ref{Al1_2} it is observed that consensus is reached rapidly, while the trajectories converge to the frequency of the mode with a relatively slow rate. This is in accord with expectations, given use of blended dynamics ideas.

\begin{figure}
\centering
\includegraphics[width=1.0\linewidth]{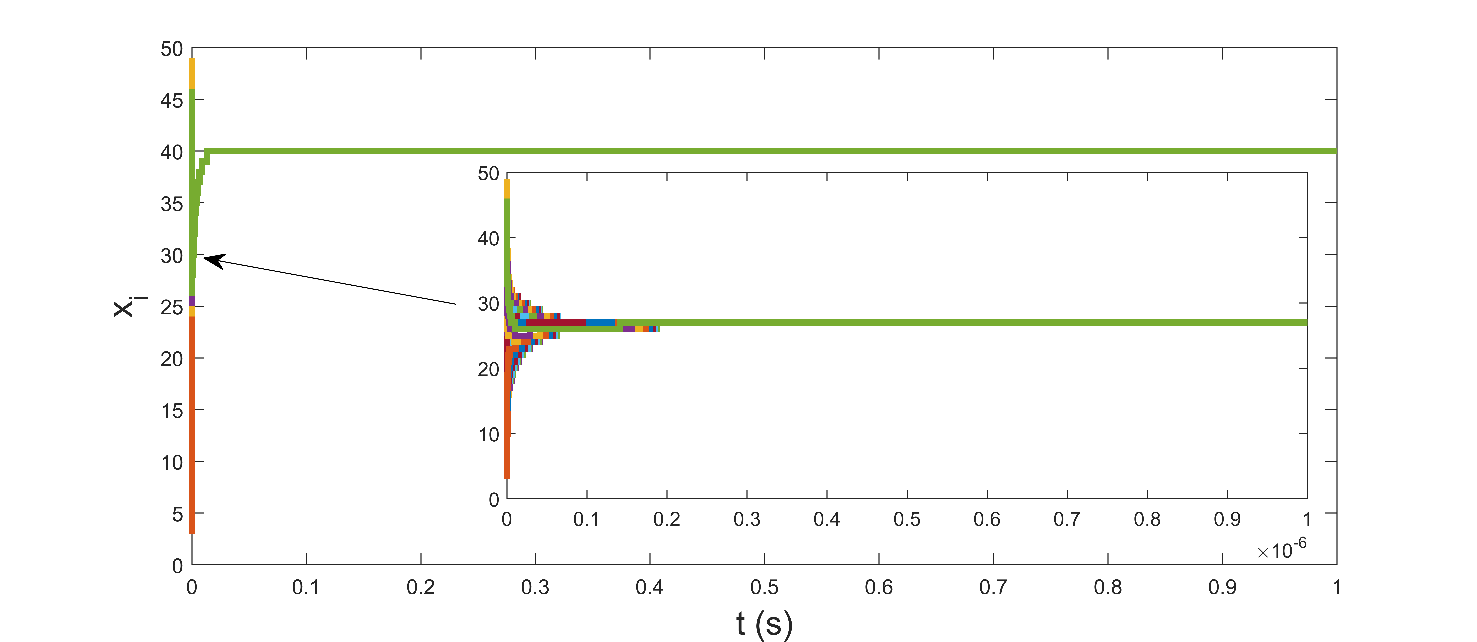}
\caption{The network size ,i.e. $N$, estimated at each agent with protocol (\ref{Nconsensus}).}
\label{fig4}
\end{figure}

\begin{figure}
\centering
\includegraphics[width=1.0\linewidth]{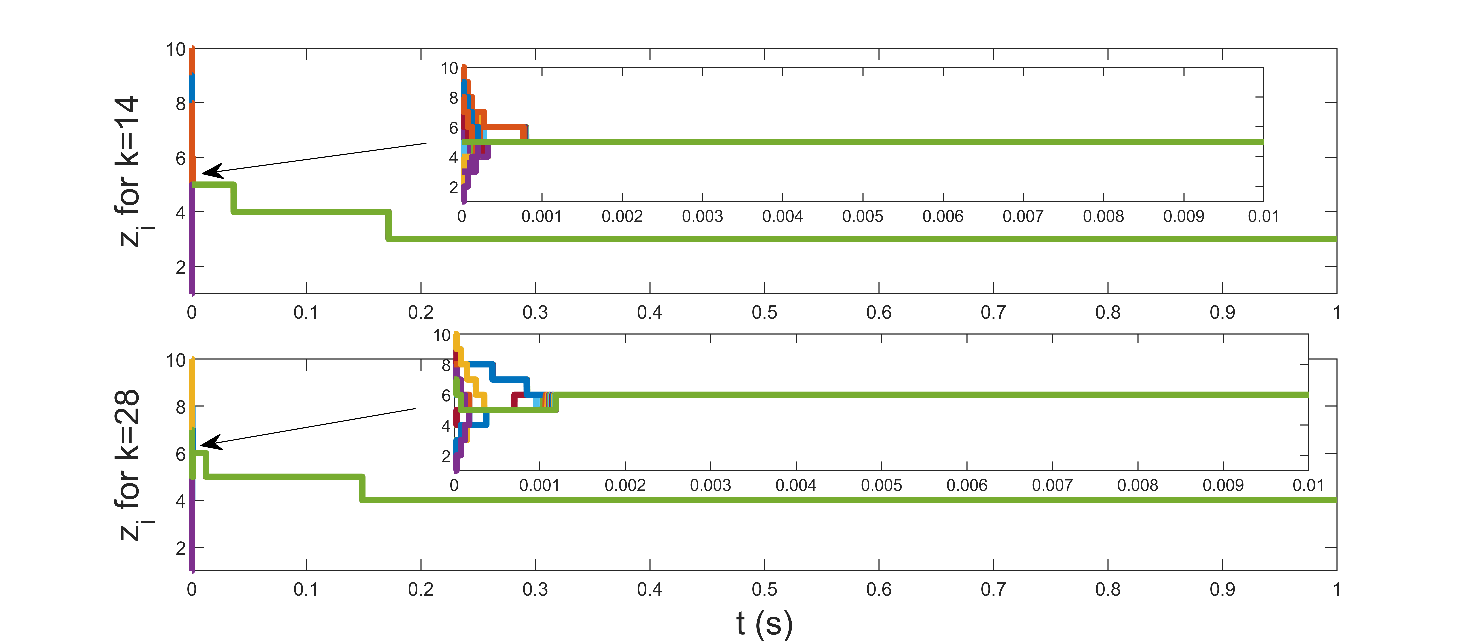}
\caption{Top figure: the estimated $14$-th smallest element at each agent with protocol (\ref{k-th1}), converging to $3$; bottom figure: the estimated $28$-th smallest element at each agent with protocol (\ref{k-th1}), converging to $4$.}
\label{fig2}
\end{figure}

\begin{figure}
\centering
\includegraphics[width=1.0\linewidth]{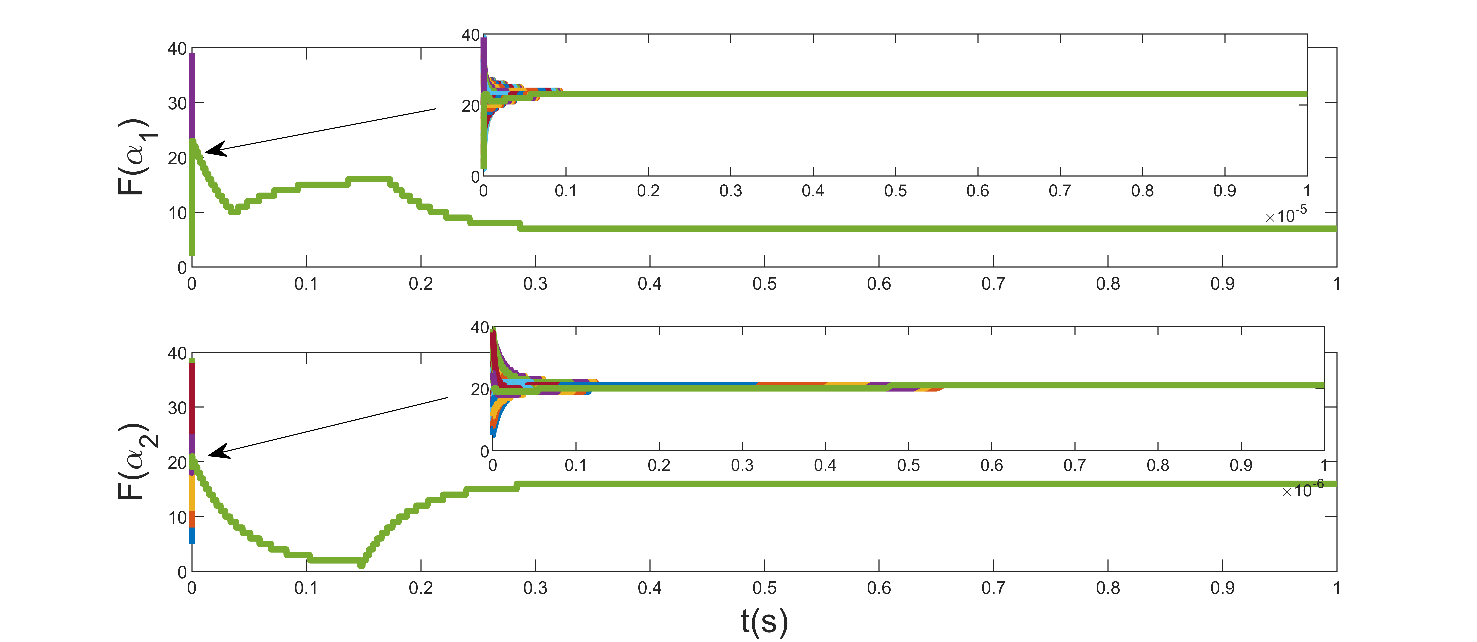}
\caption{Top figure: The estimated frequency of the $14$-th smallest element at each agent with protocol (\ref{Mode-consensus})--(\ref{ell}), converging to $7$; bottom figure: The estimated frequency of the $28$-th smallest element at each agent with protocol (\ref{Mode-consensus})--(\ref{ell}), converging to $16$;}
\label{fig3}
\end{figure}

\begin{figure}
\centering
\includegraphics[width=1.0\linewidth]{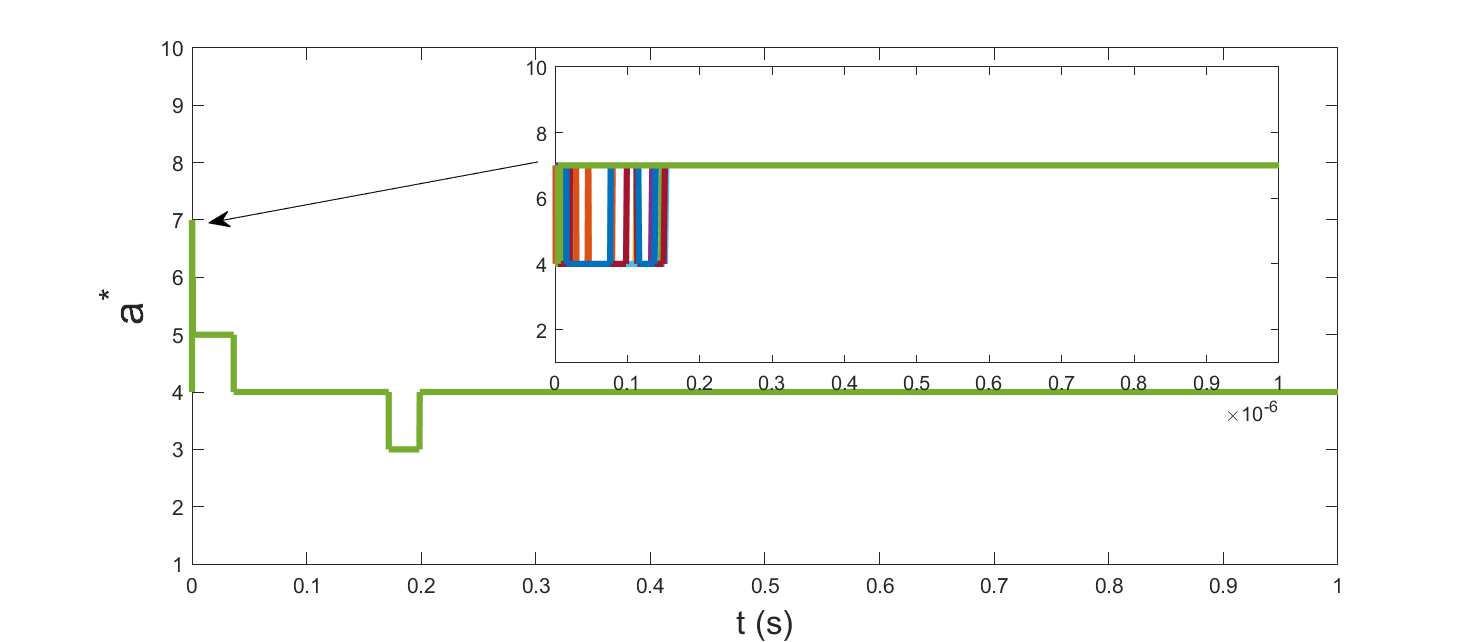}
\caption{The mode $a^*$ estimated at each agent with Algorithm \ref{Alg2}, converging to $4$.}
\label{fig1}
\end{figure}

Second, Algorithm \ref{Alg2} is examined. Suppose that $\bar N=50$. With protocol (\ref{Nconsensus}) where $h_x=10^3$ and $\gamma_x=\bar N^3=1.25\times10^5$, and the initial condition at each agent (i.e., the initial guess of $N$ at each agent) is randomly and independently selected from $1$ to $\bar N$, each agent asymptotically estimates the network size, as shown in Fig.~\ref{fig4}. 
Next, assume that $K=3$ is used in Algorithm \ref{Alg2} (this is valid since $\mathcal F\left(a^*\right)=16> \lceil{\frac{N}{K}}\rceil=14$). 
Thus only the $14$-th and $28$-th smallest element need to be estimated. 
Fig.~\ref{fig2} shows the corresponding estimation result by protocol (\ref{k-th1}), where $\beta=\frac{1}{\bar N}=0.02$,  $g=\left|\Omega\right|=10$, $\gamma_z=g\bar N^2=2.5\times10^4$ and the initial condition at each agent (i.e., the initial guess of the mode at each agent) is randomly and independently selected from $\Omega$. 
It is shown that the $14$-th element converges to $3$ while the $28$-th element converges to $4$. 
Then, Fig.~\ref{fig3} shows that the frequency of the $14$-th smallest element converges to $7$, and the $28$-th smallest element converges to $16$ by running protocol (\ref{Mode-consensus})--(\ref{ell}), where the gains and initial conditions are selected following the rules of the first simulation. Finally, Fig.~\ref{fig1} shows the mode estimated at each agent.

\begin{figure}
\centering
\includegraphics[width=1.0\linewidth]{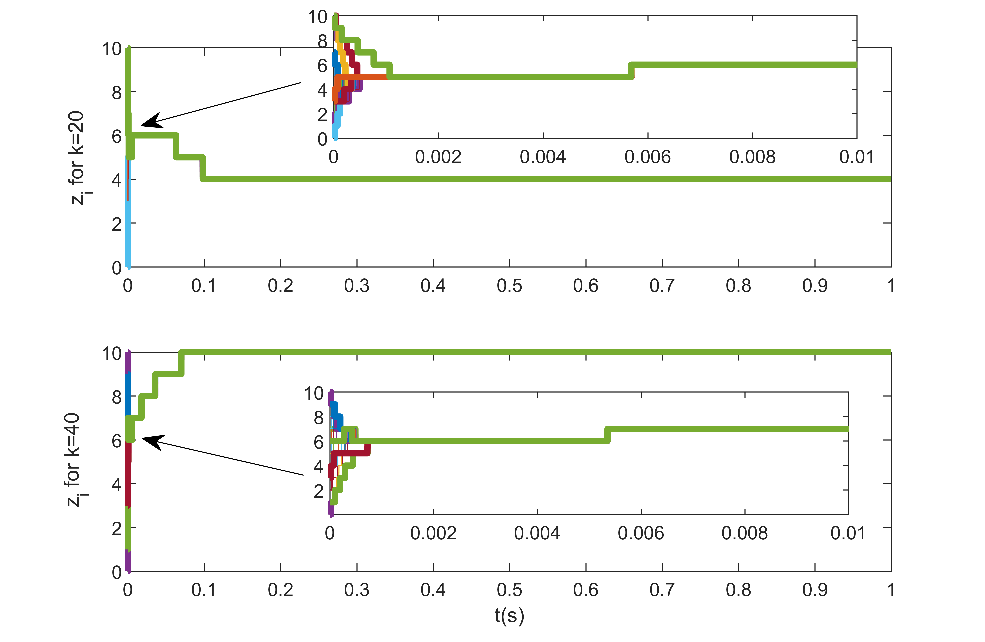}
\caption{Top figure: the estimated $20$-th smallest element at each agent with protocol (\ref{k-th1}), converging to $4$; bottom figure: the estimated $40$-th smallest element at each agent with protocol (\ref{k-th1}), converging to $10$.}.
\label{fig8}
\end{figure}

\begin{figure}
\centering
\includegraphics[width=0.8\linewidth]{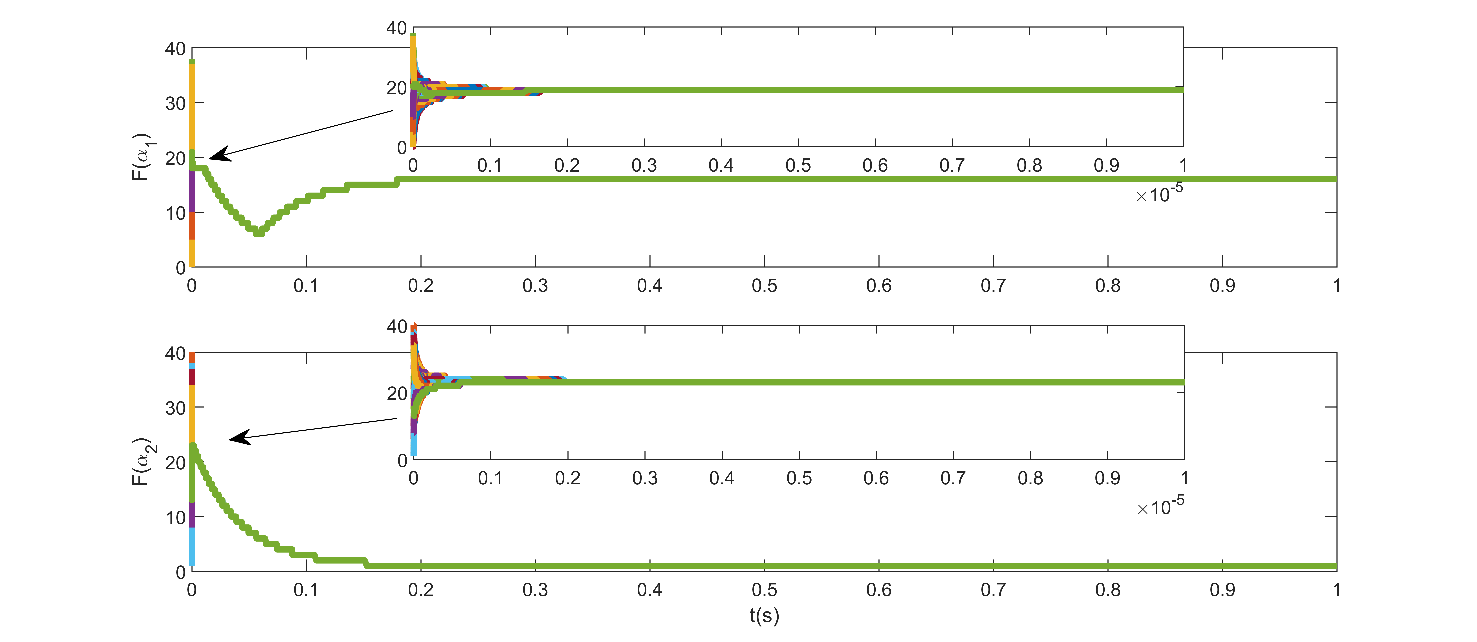}
\caption{Top figure: The estimated frequency of the $20$-th smallest element at each agent with protocol (\ref{Mode-consensus})--(\ref{ell}), converging to $4$; bottom figure: The estimated frequency of the $40$-th smallest element at each agent with protocol (\ref{Mode-consensus})--(\ref{ell}), converging to $1$.}
\label{fig7}
\end{figure}

\begin{figure}
\centering
\includegraphics[width=1.0\linewidth]{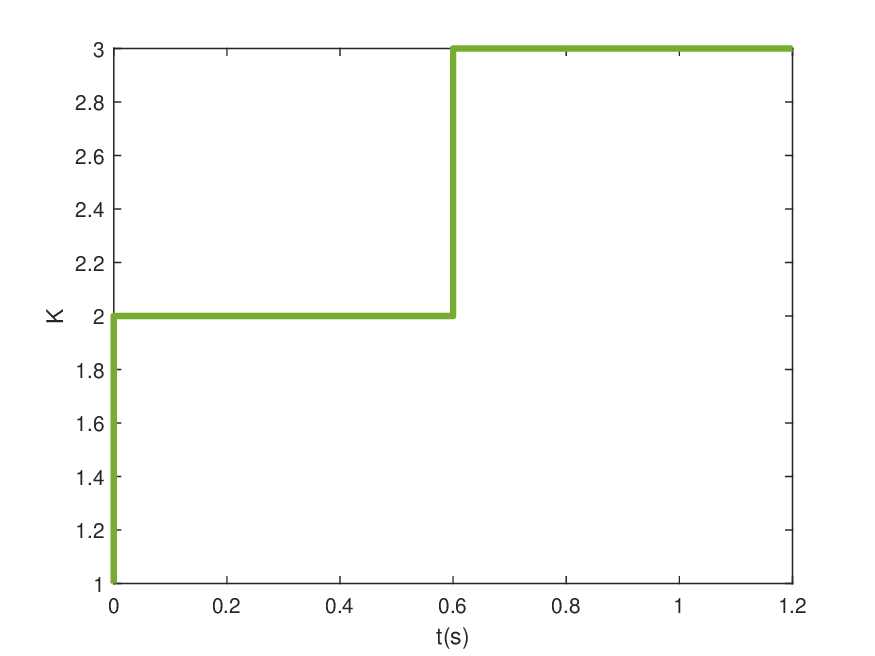}
\caption{The time evolution of $K$.}
\label{fig9}
\end{figure}

\begin{figure}
\centering
\includegraphics[width=1.0\linewidth]{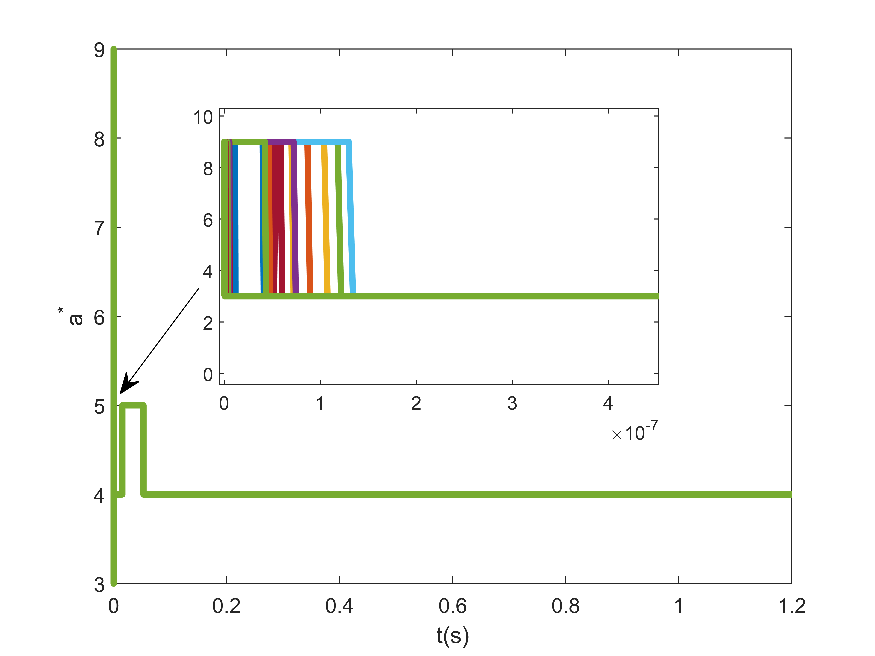}
\caption{The mode $a^*$ estimated at each agent with Algorithm \ref{Alg3}, converging to 4.}
\label{fig10}
\end{figure}

Lastly, Algorithm \ref{Alg3} is examined. Parameter settings follow Algorithms \ref{Alg1} and \ref{Alg2}, and the iteration starts at $K=1$ and $F=1$. It is supposed that the criterion ``$F<\lceil{\frac{N}{K}}\rceil$" is verified every $0.6$s. Since $F<\lceil{\frac{N}{K}}\rceil=40$, $K$ is switched to $2$ immediately, and $\lceil{\frac{N}{K}}\rceil=20$. Thus the $20$-th and $40$-th smallest element need to be estimated, and the corresponding result is shown in Fig.~\ref{fig8}-\ref{fig7}, in which the attribute estimation  converges to $4$ and $10$ and the frequency estimation converges to $16$ and $1$. Then since $F=16<\lceil{\frac{N}{K}}\rceil=20$, $K$ is further switched to $3$ in Fig. \ref{fig9}. As the situation at $K=3$ is the same as the second simulation, simulation details are omitted. Note that the termination condition is satisfied since $F=16>\lceil{\frac{N}{K}}\rceil=14$. Finally, Fig.~\ref{fig10} shows the mode estimated at each agent during the iterations.


\section{Conclusion}\label{sec:conclusion}

It is shown in this paper that, by employing a blended-dynamics based protocol, distributed mode consensus can be reached in an undirected network. It is also shown that, if the frequency of the mode is no less than $\lceil\frac{N}{K}\rceil$ for some positive integer $K$ satisfying $2K+1<\left|\Omega\right|$, the number of distinct frequencies computed at every agent can be reduced. 

To design the mode consensus algorithms, the upper bound of the network size, $\bar N$, is required
\emph{a priori} information. The convergence time is always finite.

A nice feature of the proposed algorithms is the plug-and-play property. When a new agent plugs in or leaves the network, only passive manipulations are needed for the agents in order to restore consensus to a new mode.

Future works along this direction include extending the present results to the case of directed graphs, developing second (or higher) order algorithms to reach faster convergence, adaptively adjusting the gains so that the need for knowledge of $\bar N$ can be removed.
While the piecewise-constant interaction graph is considered in this paper, gossip or other stochastic sorts of interactions are also of interest for future direction of research.



\appendix

We shall prove Theorem \ref{thm1} following the idea of \cite{lee2018distributed}. 
By letting $y = [y_1, y_2, \cdots, y_N]^T$ and $b = [ I(a,a_1), I(a,a_2), \cdots, I(a,a_N) ]^T$, the overall system \eqref{Mode-consensus}--\eqref{ell} is rewritten as
\begin{equation}\label{eq:app_y}
	\dot y(t) = h \left[-(\gamma L + e_1 e_1^T) y(t) + b \right]
\end{equation}
where $e_1=[1,0,\cdots,0]^T \in \mathbb R^N$, and $h_y$ and $\gamma_y$ in \eqref{Mode-consensus} and $\bar N$ are written as $h$, $\gamma$, and $N$, respectively, for simplicity.

Let $\lambda_2(L)$ denote the second smallest eigenvalue of the symmetric Laplacian matrix $L$, which is nonzero under Assumption \ref{Asu:connectivity}.	
Then, the following lemma is a key to the proof.

\begin{Lem}[Lemma 1 in \cite{lee2018distributed}]\label{lem:app_lem1}
	Suppose that Assumption~\ref{Asu:connectivity} holds. 
	If $\gamma > 0$, then the matrix $(\gamma L + e_1e_1^T)$ is positive definite. 
	Moreover, if $\gamma \ge N/\lambda_2(L)$, then $\lambda_{\min}\left(\gamma L + e_1e_1^T\right) \ge 1/(4N)$,
	where $\lambda_{\min}(A)$ is the smallest eigenvalue for a symmetric matrix $A$.
\end{Lem}
	
\smallskip
	
Under Lemma \ref{lem:app_lem1}, system \eqref{eq:app_y} has the exponentially stable equilibrium
$$y^* = \left(\gamma L + e_1 e_1^T\right)^{-1} b.$$
	
\begin{Lem}\label{app:lem2}
Let $y_i^*$ be the $i$-th element of $y^*$. 
Under Assumption \ref{Asu:connectivity},
\begin{enumerate}
	\item $y_1^* = 1_N^T b = \mathcal{F}(a)$,
	\item for $i=2,\dots,N$, 
	\begin{equation}\label{eq:app1}
		|y_i^* - y_1^*| < \frac{1}{\gamma} \frac{\sqrt{2} N^3}{4}.
	\end{equation}	
\end{enumerate}
\end{Lem}	

\begin{IEEEproof}
Observe that $1_N^T (\gamma L + e_1e_1^T) = e_1^T$ because $1_N^T L = 0$ and $1_N^T e_1 = 1$.
Then,
\begin{align*}
	y_1^* = e_1^T y^* = 1_N^T (\gamma L + e_1e_1^T)(\gamma L + e_1e_1^T)^{-1}b = 1_N^T b,
\end{align*}
which proves the first claim.
Under Assumption \ref{Asu:connectivity}, the matrix $L$ has exactly one zero eigenvalue and there exists an orthogonal matrix $U \in\mathbb R^{N\times N}$ such that
$$U = \begin{bmatrix} \frac{1}{\sqrt{N}} 1_N & Q \end{bmatrix} \quad \text{and} \quad
L U = U \begin{bmatrix} 0 & 0 \\ 0 & \Lambda \end{bmatrix}$$
where $Q \in \mathbb R^{N \times (N-1)}$ is orthogonal, and $\Lambda \in \mathbb R^{(N-1) \times (N-1)}$ is real and diagonal.
Then, with a coordinate change
$$\begin{bmatrix} \eta_1 \\ \tilde\eta \end{bmatrix} := U^T y = \begin{bmatrix} \frac{1}{\sqrt{N}} 1_N^T \\ Q^T \end{bmatrix} y,$$
the equilibrium $y^*$ can be expressed by
\begin{equation}\label{eq:app_ystar}
	y^* = \frac{1}{\sqrt{N}} 1_N \eta_1^* + Q \tilde\eta^*
\end{equation}
where $\eta_1^* = \lim_{t\to\infty} \eta_1(t)$ and $\tilde\eta^* = \lim_{t\to\infty} \tilde\eta(t)$ whose existence follows from the fact that $\lim_{t\to\infty} y(t) = y^*$.
In fact, we observe that
\begin{align*}
	\dot {\tilde\eta} &= Q^T \dot y = Q^T (-\gamma L y - e_1e_1^T y + b) \\
	&= -\gamma Q^T L Q \tilde\eta - Q^T e_1 y_1 + Q^T b
\end{align*}
where $Q^T L Q = \Lambda$ which is positive definite.
Therefore, we see that $\lim_{t\to\infty} \tilde\eta(t) = \tilde\eta^* = (1/\gamma) \Lambda^{-1} Q^T (b-e_1y_1^*)$.
Now, by \eqref{eq:app_ystar} and by $y_1^* = 1_N^T b$, we have that 
\begin{align*}
	e_1^T y^* &= y_1^* = \frac{1}{\sqrt{N}} \eta_1^* + \frac{1}{\gamma} e_1^T Q \Lambda^{-1} Q^T (I - e_1 1_N^T) b,
\end{align*}
from which $\eta_1^*$ is obtained.
With the expressions for $\eta_1^*$ and $\tilde\eta^*$, equation \eqref{eq:app_ystar} yields that
$$y^* = 1_N y_1^* + \frac{1}{\gamma} (I-1_Ne_1^T) Q \Lambda^{-1} Q^T (I-e_11_N^T)b.$$
Therefore, 
\begin{align*}
	y_i^* - y_1^* &= \frac{1}{\gamma} e_i^T (I-1_Ne_1^T) Q \Lambda^{-1} Q^T (I-e_11_N^T)b \\
	&= \frac{1}{\gamma} (e_i^T - e_1^T) Q \Lambda^{-1} Q^T (I-e_11_N^T)b.
\end{align*}
Noting that, for $N \ge 2$,
\begin{align*}
	\left| (I - e_11_N^T)b \right|^2 &= \left( \sum_{i=2}^N b_i \right)^2 + \sum_{i=2}^N b_i^2 \\
	&\le (N - 1) \sum_{i=2}^N b_i^2  + \sum_{i=2}^N b_i^2 = N \sum_{i=2}^N b_i^2 < N^2	
\end{align*}
we finally obtain 
\begin{align*}
	|y_i^* - y_1^*| < \frac{1}{\gamma} \cdot \sqrt{2} \cdot \frac{1}{\lambda_2(L)} \cdot N.
\end{align*}
Recalling that $\lambda_2(L) \ge 4/N^2$ \cite{mohar1991eigenvalues}, \eqref{eq:app1} follows.
\end{IEEEproof}

Now, it follows from \eqref{eq:app_y} that
$$y(t) - y^* = e^{- h (\gamma L + e_1e_1^T) t} (y(0)-y^*).$$
Here we note that, since the matrix $(\gamma L + e_1e_1^T)$ is symmetric, 
$$\|e^{- h (\gamma L + e_1e_1^T) t}\| \le k e^{-\lambda_{\min}(\gamma L + e_1e_1^T) h t} \quad \text{with $k=1$}.$$
Recalling that the assumption $\gamma \ge N^3$ of Theorem~\ref{thm1} implies $\gamma \ge N^3/4 \ge N/\lambda_2(L)$ by the fact that $\lambda_2(L) \ge 4/N^2$ \cite{mohar1991eigenvalues} so that Lemma~\ref{lem:app_lem1} guarantees $\lambda_{\min}(\gamma L + e_1e_1^T) \ge 1/(4N)$.
The assumption $\gamma \ge N^3$ also implies that $|y_i^*-y_1^*| < \sqrt{2}/4$ by \eqref{eq:app1}, with which we note that $-\sqrt{2}/4 < y_i^* < N+\sqrt{2}/4$ for all $1 \le i \le N$ because $0 \le y_1^* = {\mathcal F}(a) \le N$.
This implies that $\|y(0)-y^*\| < M_{{\mathcal K}_y}\sqrt{N}$ where $M_{{\mathcal K}_y} = N+1$ that is the size of the interval ${\mathcal K}_y = [-0.5, N+0.5]$ since $y_i(0) \in {\mathcal K}_y$ by the assumption.
Putting together, we obtain
\begin{align*}
|y_i(t) - y_1^*| &\le |y_i(t) - y_i^*| + |y_i^* - y_1^*| \\
&\le \|y(t) - y^*\| + |y_i^* - y_1^*| \\
&< e^{-\frac{h}{4N}t} \cdot  M_{{\mathcal K}_y} \sqrt{N} + \frac{\sqrt{2}}{4}.
\end{align*}
Therefore, when
$$t > \frac{4N}{h} \ln \frac{4  M_{{\mathcal K}_y} \sqrt{N}}{2-\sqrt{2}} =: {\mathcal T_y}(N),$$
we have that $|y_i(t) - {\mathcal F}(a)| < 1/2$ for all $i$.




\ifCLASSOPTIONcaptionsoff
  \newpage
\fi

\end{document}